\documentclass[twocolumn,prx,aps,amsmath,amssymb,longbibliography]{revtex4-1}

\usepackage{graphicx}
\usepackage{dcolumn}
\usepackage{bm}

\begin{document}

\title{Dual Majorana universality in thermally induced nonequilibrium}

\author{Sergey Smirnov}
\affiliation{P. N. Lebedev Physical Institute of the Russian Academy of
  Sciences, 119991 Moscow, Russia}
\email{1) sergej.physik@gmail.com\\2)
  sergey.smirnov@physik.uni-regensburg.de\\3) ssmirnov@sci.lebedev.ru}

\date{\today}

\begin{abstract}
We demonstrate that nonequilibrium nanoscopic systems with Majorana zero modes
admit special kind of universality which cannot be classified as of strictly
transport or strictly thermodynamic nature. To reveal such kind of Majorana
universality we explore purely thermal nonequilibrium states of a quantum dot
whose low-energy degrees of freedom are governed by Majorana zero
modes. Specifically, the quantum dot is coupled to a topological
superconductor, supporting Majorana zero modes, as well as to two normal
metallic contacts with the same chemical potentials but different
temperatures. It is shown that the Majorana universality in this setup is
dual: it is stored inside both the response of the electric current, excited
by exclusively the temperature difference, and the quantum dot
compressibility. The latter is defined as the derivative of the quantum dot
particle number with respect to the chemical potential and forms a universal
Majorana ratio with a proper derivative of the electric current that flows in
nonequilibrium states of purely thermal nature.
\end{abstract}

\maketitle
\section{Introduction}\label{intro}
Nanoscopic systems based on topological superconductors provide novel
low-energy degrees of freedom as key players determining their universal
physical properties. In particular, as suggested by experimental measurements
\cite{Mourik_2012,Albrecht_2016,Zhang_2018} of the differential conductance,
universality of nonequilibrium quantum transport through such systems is to a
large extent governed by low-energy quasiparticles known as Majorana zero
modes. These non-Abelian modes are often termed as Majorana fermions since
they represent their own antiquasiparticles as it happens for Abelian Majorana
fermions \cite{Majorana_1937} in quantum field theory
\cite{Itzykson_Zuber_1980}. In the context of condensed matter physics
Majorana zero modes are predicted to arise in the topological phase of the
Kitaev tight-binding chain model \cite{Kitaev_2001}. The latter has become of
extreme practical importance due to its various appealing mappings
\cite{Alicea_2012,Flensberg_2012,Sato_2016,Aguado_2017,Lutchyn_2018} onto
experimentally feasible models such as setups combining superconductors with
topological insulators \cite{Fu_2008,Fu_2009} and semiconductors whose
low-energy spectra result from an interplay between spin-orbit interactions
and an induced superconducting order parameter \cite{Lutchyn_2010,Oreg_2010}.

An alternative to transport experiments measuring mean quantities is to access
Majorana universality via fluctuations of transport quantities. For example,
noise of the electric current offers unique Majorana behavior both in the
static limit \cite{Liu_2015,Liu_2015a,Beenakker_2015,Haim_2015} and at finite
frequencies \cite{Valentini_2016,Bathellier_2019}. In particular, fluctuation
universality of Majorana zero modes is revealed in zero frequency noise via
universal effective charges \cite{Smirnov_2017} and in finite frequency
quantum noise via universal plateaus, resonances and antiresonances located at
specific frequencies \cite{Smirnov_2019}. Moreover, current shot noise allows
one to reveal also the nonlocality of Majorana zero modes
\cite{Manousakis_2020}.

Advanced experiments \cite{Hartman_2018} on nanoscopic systems have set the
stage for a very original access to Majorana universality via thermodynamic
measurements as proposed in Ref. \cite{Sela_2019}. This approach is
fundamentally different from quantum transport experiments and allows one to
describe thermodynamics of the Kitaev's chain, in particular, its Majorana
universal fractional entropy \cite{Smirnov_2015}.

Setups where nonequilibrium is simultaneously induced by a bias voltage $V$
and a temperature difference $\Delta T$ uncover even more unique physics of
Majorana zero modes which manifests, {\it e.g.}, in a violation of the
Wiedemann-Franz law \cite{Ramos-Andrade_2016} and in thermoelectric
fluctuations having a high degree of universality. Indeed, in nanoscopic
Majorana setups not only linear but also nonlinear zero frequency
thermoelectric noise turns out to be universal \cite{Smirnov_2018} while
finite frequency thermoelectric quantum noise reveals universal symmetry and
dynamic resonances with universal maxima \cite{Smirnov_2019a}. Recently, it
has been proposed that one obtains the entropy of a nanoscopic system from
thermoelectric transport experiments \cite{Kleeorin_2019} which incorporate
measurements of the differential conductance of the nanoscopic system and
measurements of its thermopower. This shows that thermoelectric transport may
also become an effective experimental tool able to detect the Majorana
universal fractional entropy \cite{Smirnov_2015} of the Kitaev's chain.

In this paper we focus on nonequilibrium states of purely thermal nature that
is induced solely by a temperature difference $\Delta T$ assuming
$V=0$. Remarkably, such kind of nonequilibrium reveals existence of Majorana
universality which is classified as neither strictly transport nor strictly
thermodynamic, {\it i.e.} none of the only two known in Majorana
experiments. We show that combining transport measurements of the electric
current in a quantum dot with measurements of the quantum dot compressibility
allows one to access such type of Majorana universality. Although
compressibility itself may be used to probe quantum phase transitions leading
to formation of Majorana zero modes \cite{Nozadze_2016}, it has never been
employed in conjunction with low-energy quantum transport governed by
non-Abelian Majorana quasiparticles.

The paper is organized as follows. In Sec. \ref{model} we present a
theoretical model of a nanoscopic system where Majorana zero modes are
involved in nonequilibrium states of purely thermal
nature. Sec. \ref{th_ind_curr} shows that the current induced in such
nonequilibrium states does not allow one to access Majorana universal
behavior. We demonstrate in Sec. \ref{M_univ_qd_compress} that in order to
reveal Majorana universality in purely thermal nonequilibrium, it is necessary
to combine the transport response stored in the induced current with the
thermodynamic response stored in the quantum dot compressibility. Finally,
with Sec. \ref{concl} we conclude the paper.
\section{Majorana nanoscopic setup in thermal nonequilibrium}\label{model}
\begin{figure}
\includegraphics[width=8.0 cm]{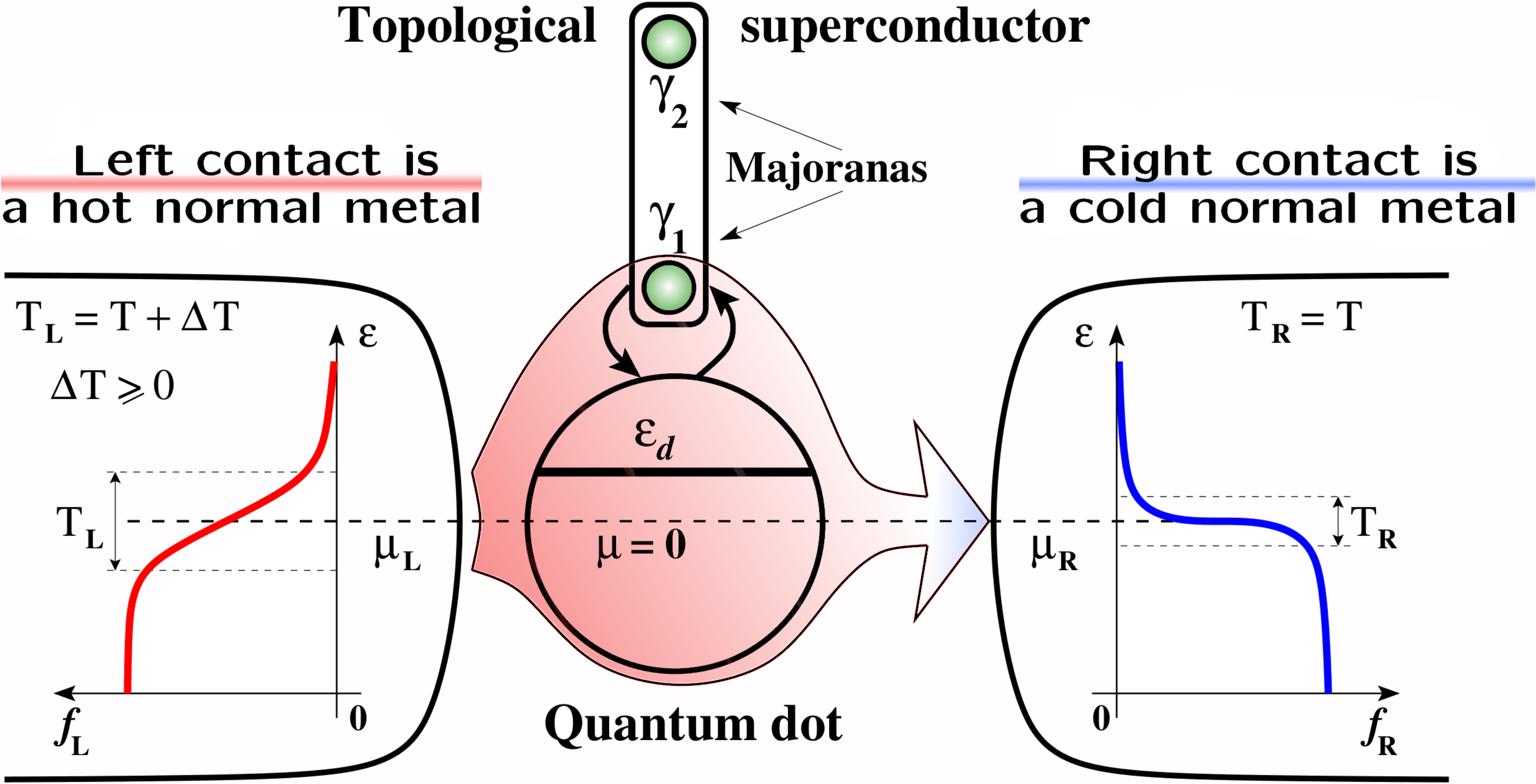}
\caption{\label{figure_1} A basic outline of a nanoscopic physical system
  where nonequilibrium quantum transport may experimentally be implemented by
  purely thermal means. The system is composed of a quantum dot interacting
  via tunneling mechanisms with left ($L$) and right ($R$) contacts as well as
  with a grounded one-dimensional topological superconductor hosting two
  Majorana zero modes at its ends, specified as $\gamma_1$ and $\gamma_2$
  (green circles), of which the first one interacts with the quantum dot (arc
  arrows). In particular, there is no any bias voltage in this system,
  {\it i.e.} the chemical potentials $\mu_{L}$ and $\mu_R$ of the left and
  right contacts, respectively, coincide, $\mu_L=\mu_R=\mu=0$. The
  temperatures of the left and right contacts are, respectively,
  $T_L=T+\Delta T$ and $T_R=T$ with the temperature difference
  $\Delta T\geqslant 0$ and the temperature $T\geqslant 0$.}
\end{figure}

To reveal a special type of Majorana universality which cannot be accessed via
pure transport or pure thermodynamic experiments it is enough to resort to the
simple model which is schematically shown in Fig. \ref{figure_1}. It includes
a quantum dot with a nondegenerate single-particle energy level $\epsilon_d$
as measured with reference to the chemical potential $\mu$. The left and right
normal metals represent the contacts which are linked to the quantum
dot. These links provide a quasiparticle exchange between the quantum dot and
contacts via quantum mechanical tunneling. The one-dimensional topological
superconductor is grounded and supports a pair of Majorana zero modes at its
ends. One of these ends is linked to the quantum dot. This link establishes a
special channel for Majorana tunneling between the quantum dot and topological
superconductor. We assume that the left and right contacts are described by
equilibrium Fermi-Dirac distributions with the chemical potentials $\mu_{L,R}$
and temperatures $T_{L,R}$,
\begin{equation}
f_{L,R}(\epsilon)=\biggl[\exp\bigl(\frac{\epsilon-\mu_{L,R}}{k_\text{B}T_{L,R}}\bigl)+1\biggl]^{-1}.
\label{FD_distributions}
\end{equation}
Since no bias voltage is applied, $V=0$, we have
$\mu_L=\mu_R=\mu=0$. Nonequilibrium states arise only due to finite $\Delta T$
or thermal voltage $eV_T\equiv k_\text{B}\Delta T$.

For quantitative analysis of the quantum transport in the above system we
represent its Hamiltonian as the sum
$\hat{H}=\hat{H}_d +\hat{H}_c +\hat{H}_{ts} +\hat{H}_{d-c} +\hat{H}_{d-ts}$.
The Hamiltonians of the quantum dot, contacts and topological superconductor
are $\hat{H}_d=\epsilon_d d^\dagger d$,
$\hat{H}_c=\sum_{l=L,R}\sum_k\epsilon_k c_{lk}^\dagger c_{lk}$ and
$\hat{H}_{ts}=i\xi\gamma_2\gamma_1/2$. The tunneling Hamiltonians,
$\hat{H}_{d-c}=\sum_{l=L,R}\sum_k T_{lk}c_{lk}^\dagger d+\text{H.c.}$ and
$\hat{H}_{d-ts}=\eta^* d^\dagger\gamma_1+\text{H.c.}$, describe, respectively,
the interactions of the quantum dot with the contacts and topological
superconductor. The contacts are massive normal metals with a continuous
energy spectrum $\epsilon_k$. For simplicity, we use the traditional
approximation assuming that the contacts density of states $\nu(\epsilon)$
varies sufficiently weakly over the energies involved in the quantum
transport, that is in fact energy independent,
$\nu(\epsilon)\approx\nu_c/2$. The Majorana operators $\gamma_{1,2}$ are
self-adjoint, $\gamma^\dagger_{1,2}=\gamma_{1,2}$, and their anticommutator is
$\{\gamma_i,\gamma_j\}=2\delta_{ij}$. The energy $\xi$ characterizes the
overlap of the Majorana modes so that perfectly separated Majoranas correspond
to $\xi=0$. Another conventional assumption is that the tunneling between the
quantum dot and contacts is independent of the quantum numbers $l,k$, that is
$T_{lk}=\mathcal{T}$. Then the coupling of the quantum dot and the left/right
contact is expressed in terms of the quantity
$\Gamma=2\pi\nu_c|\mathcal{T}|^2$. The coupling between the quantum dot and
topological superconductor is specified by the quantity $|\eta|$.

The quasiparticle current, induced by $\Delta T$, may be derived by means of
the Keldysh field integral \cite{Altland_2010} written in terms of the Keldysh
action $S_K$ with sources $J_l(t)$,
\begin{equation}
\begin{split}
&Z[J_l(t)]=\int\mathcal{D}[\theta(t)]e^{\frac{i}{\hbar}S_K[\theta(t);J_l(t)]},\\
&S_K=S_{sys}+S_{scr},
\end{split}
\label{GenFunc}
\end{equation}
where $\{\theta(t)\}=\{\psi(t),\phi_{lk}(t),\zeta(t)\}$ are the Grassmann
fields of the quantum dot, $\psi(t)$, contacts, $\phi_{lk}(t)$, and
topological superconductor, $\zeta(t)$, defined on the forward ($q=+$) and
backward ($q=-$) branches of the Keldysh contour, $S_{sys}$ is the system
action having the conventional form \cite{Smirnov_2019} in the
retarded-advanced space and $S_{scr}$ is the source action involving the
current operator,
\begin{equation}
\begin{split}
&S_{scr}[J_l(t)]=-\int_{-\infty}^\infty dt\sum_{l=L,R}\sum_{q=+,-}J_{lq}(t)\hat{I}_{lq}(t),\\
&\hat{I}_{lq}(t)=\frac{ie}{\hbar}\sum_k[\mathcal{T}\bar{\phi}_{lkq}(t)\psi_q(t)-\mathcal{T}^*\bar{\psi}_q(t)\phi_{lkq}(t)].
\end{split}
\label{CurrOpr}
\end{equation}

The current in contact $l$ is obtained via the functional derivative over the
source field taken at $J_{lq}(t)=0$ and arbitrary $q$,
\begin{equation}
I_l=\langle\hat{I}_{lq}(t)\rangle=i\hbar\frac{\delta Z[J_l(t)]}{\delta J_{lq}(t)}\biggl|_{J_{lq}(t)=0}.
\label{CurrFuncDerv}
\end{equation}
Below we focus on the quasiparticle current $I$ in the hot, {\it i.e.} left,
contact, $I\equiv I_L$, as a function of $V_T$, $\epsilon_d$ and
$|\eta|$. From Eq. (\ref{CurrFuncDerv}) one finds the following expression
\cite{Meir_1992},
\begin{equation}
\begin{split}
&I(V_T,\epsilon_d,|\eta|)=\\
&=-\frac{e\Gamma}{\hbar^2}\int_{-\infty}^\infty\frac{d\epsilon}{2\pi}\bigl\{\text{Im}[G^R(\epsilon)]f_L(\epsilon)-\frac{i}{2}G^<(\epsilon)\bigl\},
\end{split}
\label{Meir_Wingreen}
\end{equation}
where the quantum dot retarded/lesser Green's functions, $G^{R/<}$, follow
from the inverse kernel of $S_{sys}$.
\begin{figure}
\includegraphics[width=8.0 cm]{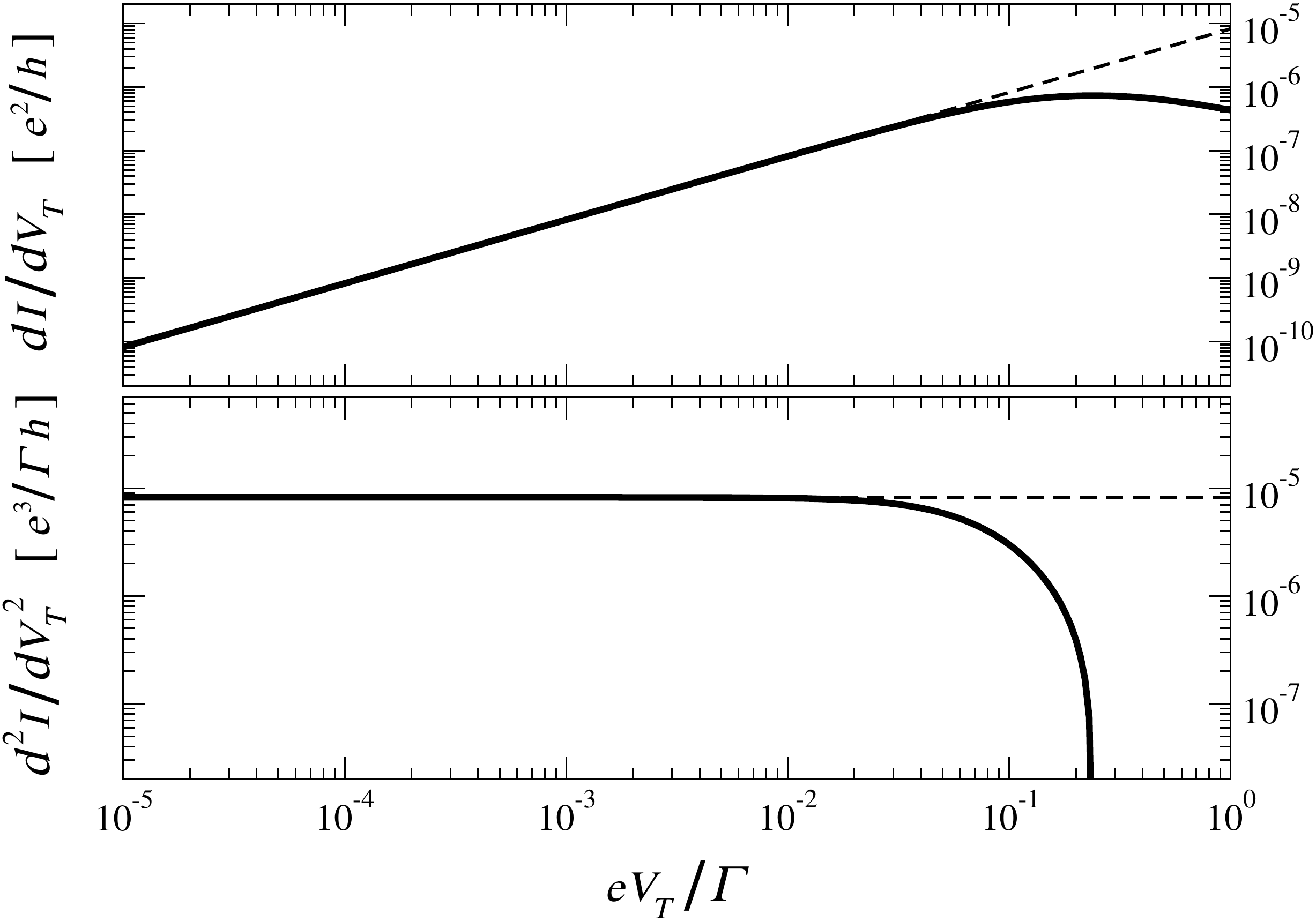}
\caption{\label{figure_2} The first and the second derivatives of the electric
  current with respect to the thermal voltage,
  $\partial I(V_T,\epsilon_d,|\eta|)/\partial V_T$,
  $\partial^2 I(V_T,\epsilon_d,|\eta|)/\partial V_T^2$, are shown as functions
  of the thermal voltage $V_T$. Here $\epsilon_d/\Gamma=10$,
  $k_\text{B}T/\Gamma=10^{-12}$, $|\eta|/\Gamma=10^3$,
  $\xi/\Gamma=10^{-8}$.}
\end{figure}
\section{Thermally induced Majorana current}\label{th_ind_curr}
In Fig. \ref{figure_2} we show the results obtained for the first and the
second derivatives of the electric current with respect to the thermal
voltage. At small values of the thermal voltage $eV_T$ the first derivative
(upper panel) has a linear dependence on $eV_T$. As a consequence, at small
values of $eV_T$ the second derivative (lower panel) is independent of the
thermal voltage. As one can see, it saturates to a certain value, the
coefficient in the linear dependence of the first derivative
$\partial I(V_T,\epsilon_d,|\eta|)/\partial V_T$ on $eV_T$. This coefficient
is a function of the quantum dot energy level $\epsilon_d$ (or, equivalently,
the chemical potential $\mu$) controlled by a gate voltage as well as a
function of the energy $|\eta|$ characterizing the strength of the Majorana
tunneling between the quantum dot and topological superconductor.

Note, that although for the present setup it is not crucial whether
$\epsilon_d>0$ or $\epsilon_d<0$, we nevertheless prefer to use positive
values of $\epsilon_d$ to disentangle Majorana universality from Kondo
universality in more general interacting setups
\cite{Silva_2020,Weymann_2020}. As it is known \cite{Hewson_1997}, the Kondo
universality may arise when $\epsilon_d<0$.

\begin{figure}
\includegraphics[width=8.0 cm]{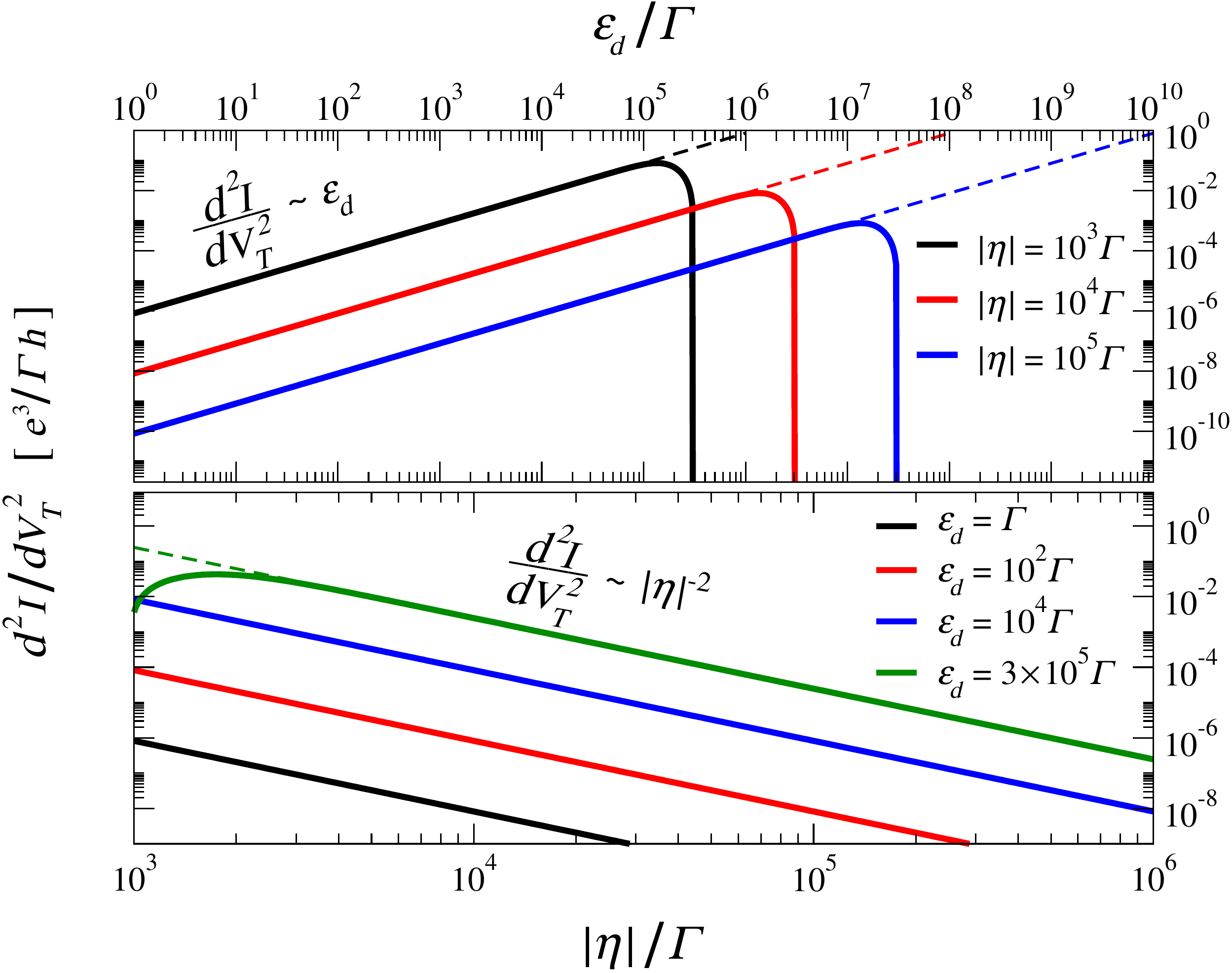}
\caption{\label{figure_3} The second derivative
  $\partial^2 I(V_T,\epsilon_d,|\eta|)/\partial V_T^2$ is shown for a small
  value of the thermal voltage, specifically, for $eV_T/\Gamma=10^{-5}$. The
  other parameters have the same values as in Fig. \ref{figure_2}. }
\end{figure}

The behavior of the second derivative
$\partial^2 I(V_T,\epsilon_d,|\eta|)/\partial V_T^2$ as a function of
$\epsilon_d$ and $|\eta|$ is analyzed in Fig. \ref{figure_3} at small
$eV_T$. In the upper panel
$\partial^2 I(V_T,\epsilon_d,|\eta|)/\partial V_T^2$ is shown for various
values of the energy $|\eta|$ as a function of the quantum dot energy level
$\epsilon_d$ (or, equivalently, the chemical potential $\mu$) which may be
controlled by a gate voltage. The curves exhibit a linear dependence of the
second derivative $\partial^2 I(V_T,\epsilon_d,|\eta|)/\partial V_T^2$ on the
energy level $\epsilon_d$ up to $\epsilon_d\approx 10^2|\eta|$. In the lower
panel $\partial^2 I(V_T,\epsilon_d,|\eta|)/\partial V_T^2$ is shown for
various values of $\epsilon_d$ as a function of the Majorana tunneling
strength $|\eta|$. The curves exhibit an inverse quadratic dependence of the
second derivative $\partial^2 I(V_T,\epsilon_d,|\eta|)/\partial V_T^2$ on the
energy $|\eta|$ down to $|\eta|\approx 10^{-2}\epsilon_d$.

Using the numerical approach of Ref. \cite{Smirnov_2018} for obtaining
asymptotics, we find for $\xi\ll eV_T\ll\Gamma$ the following asymptotic
limit:
\begin{equation}
\frac{\partial I(V_T,\epsilon_d,|\eta|)}{\partial V_T}=\frac{e^2}{h}\frac{\pi^2}{12}\frac{\epsilon_d\,(eV_T)}{|\eta|^2}.
\label{DI_DVT}
\end{equation}
As in Ref. \cite{Smirnov_2018}, the analytical expression in
Eq. (\ref{DI_DVT}) is obtained by inspection of numerical results which
reproduce Eq. (\ref{DI_DVT}) with any desired numerical precision when the
corresponding inequalities are satisfied as strong as necessary for that
precision.

The parameters which can be varied in an experiment are $V_T$ and
$\epsilon_d$. Independence of these parameters is achieved via measurements of
the derivative
$\partial^3 I(V_T,\epsilon_d,|\eta|)/\partial\epsilon_d\partial V_T^2=e^3\pi^2/12h|\eta|^2$.
The first derivative of the current with respect to the thermal voltage may be
interpreted as a special kind of conductance which measures the current
sensitivity to the temperature inhomogeneity $\Delta T$. One can call it
thermoelectric conductance. The second derivative of the current with respect
to the thermal voltage is therefore the first derivative of the thermoelectric
conductance. It shows how the thermoelectric conductance of the quantum dot is
enhanced or suppressed by temperature inhomogeneities of the external
environment whose role is played here by the massive metallic contacts. In
contrast, the third derivative of the current with respect to the chemical
potential probes internal sensitivity of the tunneling density of states of
the quantum dot when the energy level $\epsilon_d$ is varied by a gate
voltage. Therefore, the derivative
$\partial^3 I(V_T,\epsilon_d,|\eta|)/\partial\epsilon_d\partial V_T^2$
is a very comprehensive physical quantity: it measures the transport response
to the temperature inhomogeneity of the external environment and at the same
time it provides the internal spectral response of the system.

However, the dependence of this derivative on the parameter $|\eta|$ still
remains. This parameter is not directly controlled in an experiment and will
have in general different values for different experimental setups. As a
result, the strictly transport response does not allow one to obtain universal
properties of the Majorana zero modes in purely thermal nonequilibrium states
induced by $\Delta T$ at $V=0$.
\begin{figure}
\includegraphics[width=8.0 cm]{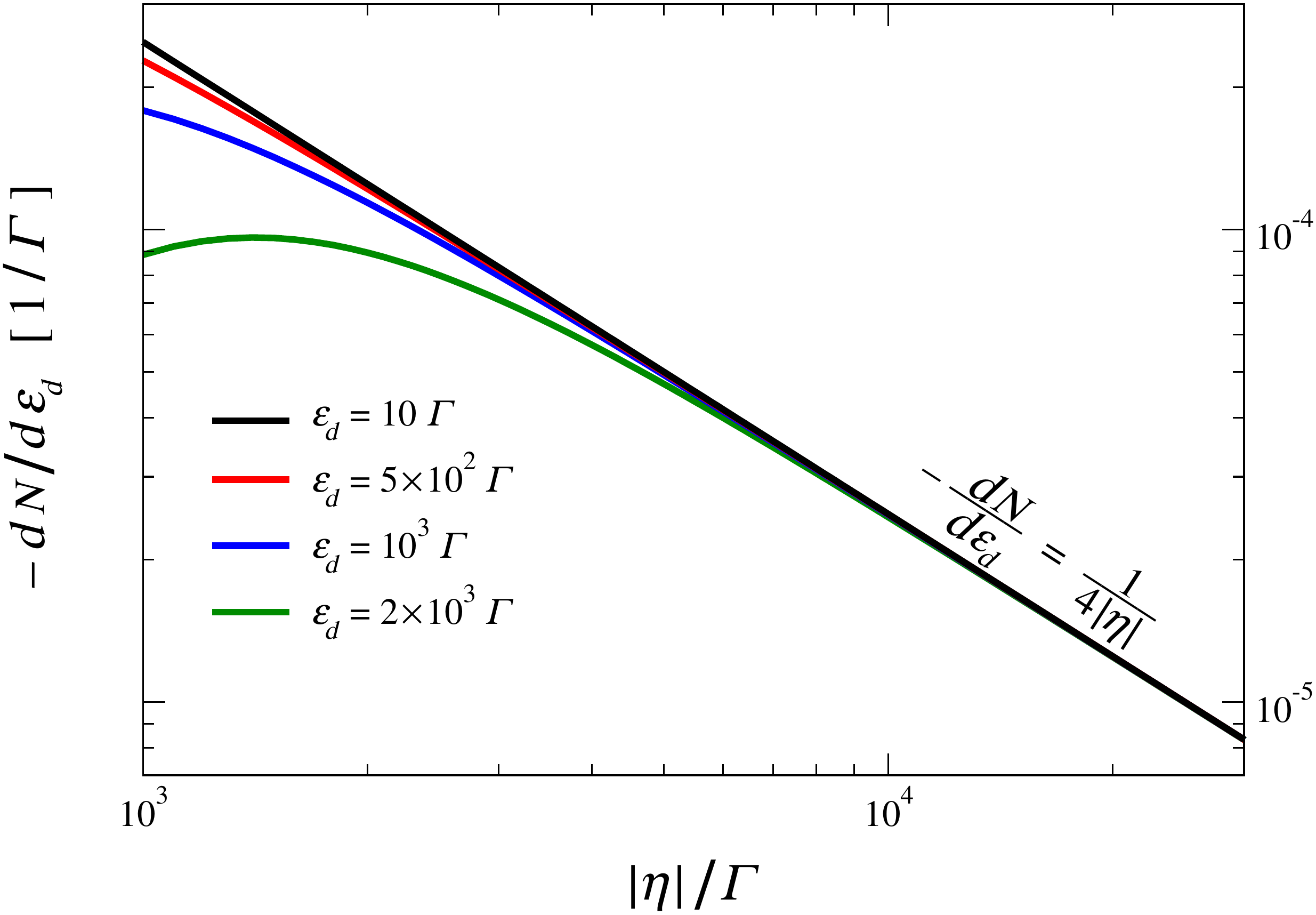}
\caption{\label{figure_4} The quantum dot compressibility, that is the first
  derivative of the quantum dot particle number with respect to the chemical
  potential (or, equivalently, the quantum dot energy level),
  $\partial N(V_T,\epsilon_d,|\eta|)/\partial\mu=-\partial N(V_T,\epsilon_d,|\eta|)/\partial\epsilon_d$.
  For all the curves the thermal voltage is small,
  $eV_T/\Gamma=10^{-5}$. The other parameters have the same values as in
  Fig. \ref{figure_2}.}
\end{figure}
\section{Majorana universality via the quantum dot compressibility}\label{M_univ_qd_compress}
Nevertheless, it is possible to uncover the Majorana universality encoded in
this purely thermal nonequilibrium if in addition one turns attention to
thermodynamic properties, namely to the quantum dot compressibility defined as
$\partial N/\partial\mu$ or, equivalently, $-\partial N/\partial\epsilon_d$,
where $N$ is the quantum dot particle number,
\begin{equation}
N(V_T,\epsilon_d,|\eta|)=-\int_{-\infty}^\infty\frac{d\epsilon}{2\pi i\hbar}G^<(\epsilon).
\label{QD_PN}
\end{equation}

In Fig. \ref{figure_4} the quantum dot compressibility is shown as a function
of the energy $|\eta|$ characterizing the strength of the Majorana tunneling
between the quantum dot and topological superconductor. Various curves
correspond to various values of the quantum dot energy level $\epsilon_d$ (or,
equivalently, the chemical potential $\mu$) controlled by a gate voltage. The
curves demonstrate that down to $|\eta|\approx 3\epsilon_d$ the
compressibility has an inverse dependence on the energy $|\eta|$. More
exactly, similar to Eq. (\ref{DI_DVT}), we find for $\xi\ll eV_T\ll\Gamma$,
the following asymptotic limit:
\begin{equation}
-\frac{\partial N(V_T,\epsilon_d,|\eta|)}{\partial\epsilon_d}=\frac{1}{4|\eta|}.
\label{QD_Compress}
\end{equation}
\begin{figure}
\includegraphics[width=8.0 cm]{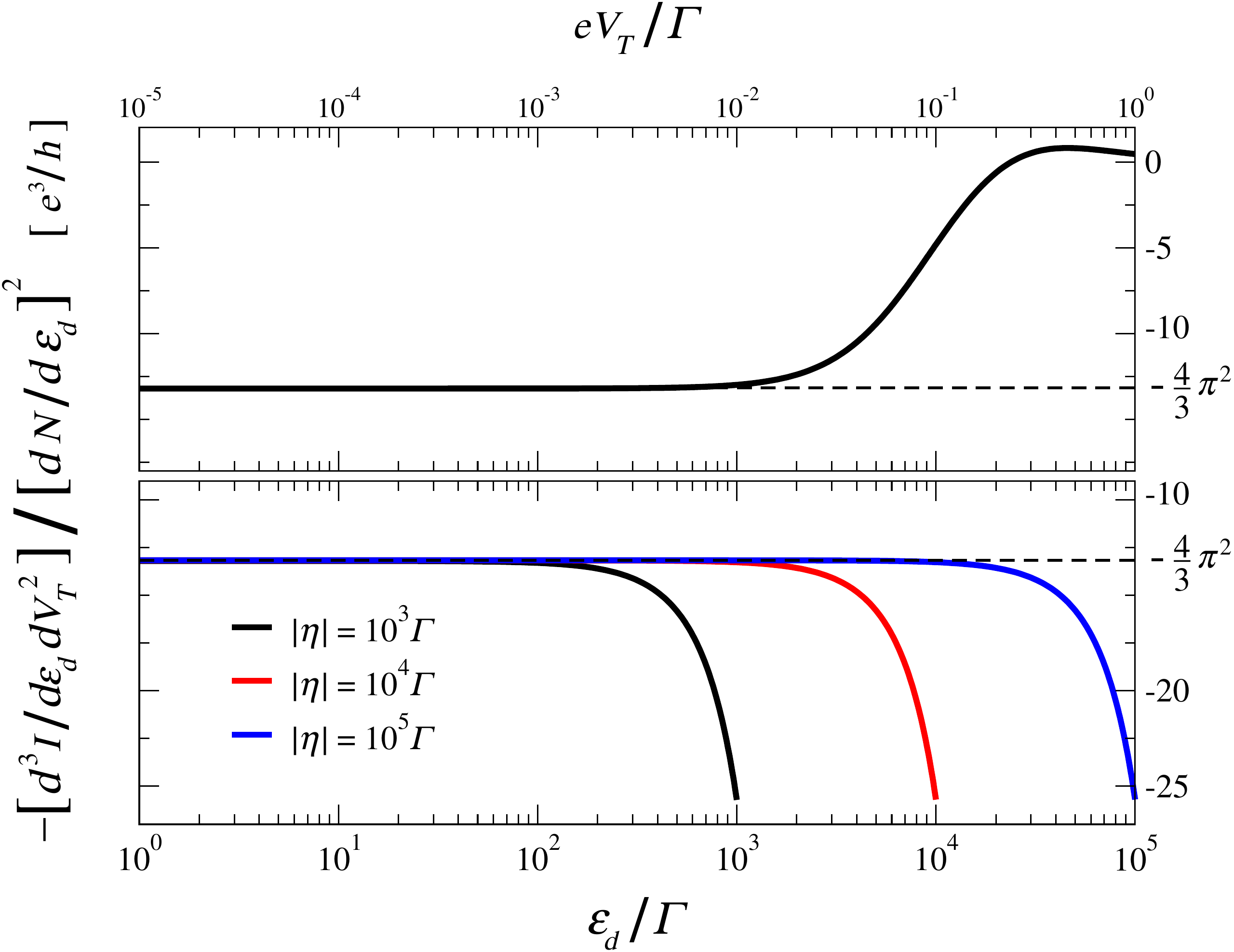}
\caption{\label{figure_5} The ratio between the derivative of the electric
  current
  $\partial^3 I/\partial\mu\partial V_T^2=-\partial^3I/\partial\epsilon_d\partial V_T^2$
  and the square of the quantum dot compressibility
  $[\partial N/\partial\mu]^2=[\partial N/\partial\epsilon_d]^2$. Upper panel:
  the ratio is shown as a function of the thermal voltage $V_T$. Lower panel:
  the ratio is shown as a function of the quantum dot energy level
  $\epsilon_d$ for various values of the Majorana tunneling strength
  $|\eta|$.}
\end{figure}

From Eqs. (\ref{DI_DVT}) and (\ref{QD_Compress}) there follows the dual (that
is having both transport and thermodynamic nature) universal ratio:
\begin{equation}
-\frac{\partial^3 I(V_T,\epsilon_d,|\eta|)/\partial\epsilon_d\partial V_T^2}{[\partial N(V_T,\epsilon_d,|\eta|)/\partial\epsilon_d]^2}
=-\frac{4\pi^2}{3}\frac{e^3}{h}.
\label{Univ_Ratio}
\end{equation}

In the upper panel of Fig. \ref{figure_5} this ratio is shown as a function of
the thermal voltage $eV_T$ with all the other parameters having the same
values as in Fig. \ref{figure_2}. The curve demonstrates that the ratio does
not depend on the thermal voltage when $eV_T/\Gamma\lesssim 10^{-2}$. In this
regime the ratio reaches the universal Majorana value $-4\pi^2e^3/3h$. In the
lower panel of Fig. \ref{figure_5} this ratio is shown for a small value of
the thermal voltage, specifically, for $eV_T/\Gamma=10^{-5}$, as a function of
the quantum dot energy level $\epsilon_d$ which may be controlled by a gate
voltage. The other parameters have the same values as in
Fig. \ref{figure_2}. The curves demonstrate that as soon as
$\epsilon_d\lesssim 10^{-1}|\eta|$, the ratio does not depend on both the
quantum dot energy level $\epsilon_d$ and the energy $|\eta|$ which
characterizes the strength of the Majorana tunneling. In this regime the ratio
reaches the universal Majorana value $-4\pi^2e^3/3h$.

When the two Majorana modes start to significantly overlap, they form a
partially separated Andreev bound state. In many respects this situation may
be adequately analyzed using the energy $\xi$ while for a more complete
analysis a measure for spatial separation \cite{Deng_2018} of the two Majorana
modes could be introduced. Here for simplicity we focus on purely energetic
arguments. Thus, when $\xi$ is large enough, our model describes a quantum dot
coupled to one end of a topological superconductor supporting a partially
separated Andreev bound state localized at that end \cite{Hell_2018}. Below we
assume weak overlaps of the Majorana zero modes. This means that in the
partially separated Andreev bound state, composed of $\gamma_1$ and
$\gamma_2$, the second Majorana mode $\gamma_2$ is still far enough so that
the quantum dot does not directly couple to $\gamma_2$. A more complicated
analysis of a model with additional direct coupling between the quantum dot
and $\gamma_2$ will be performed in another work. In pure electric transport
($\Delta T=0$) the differential conductance ($G=\partial I/\partial V$) may
behave similar \cite{Moore_2018} for both Majorana zero modes and partially
separated Andreev bound states. It is thus
\begin{figure}
\includegraphics[width=8.0 cm]{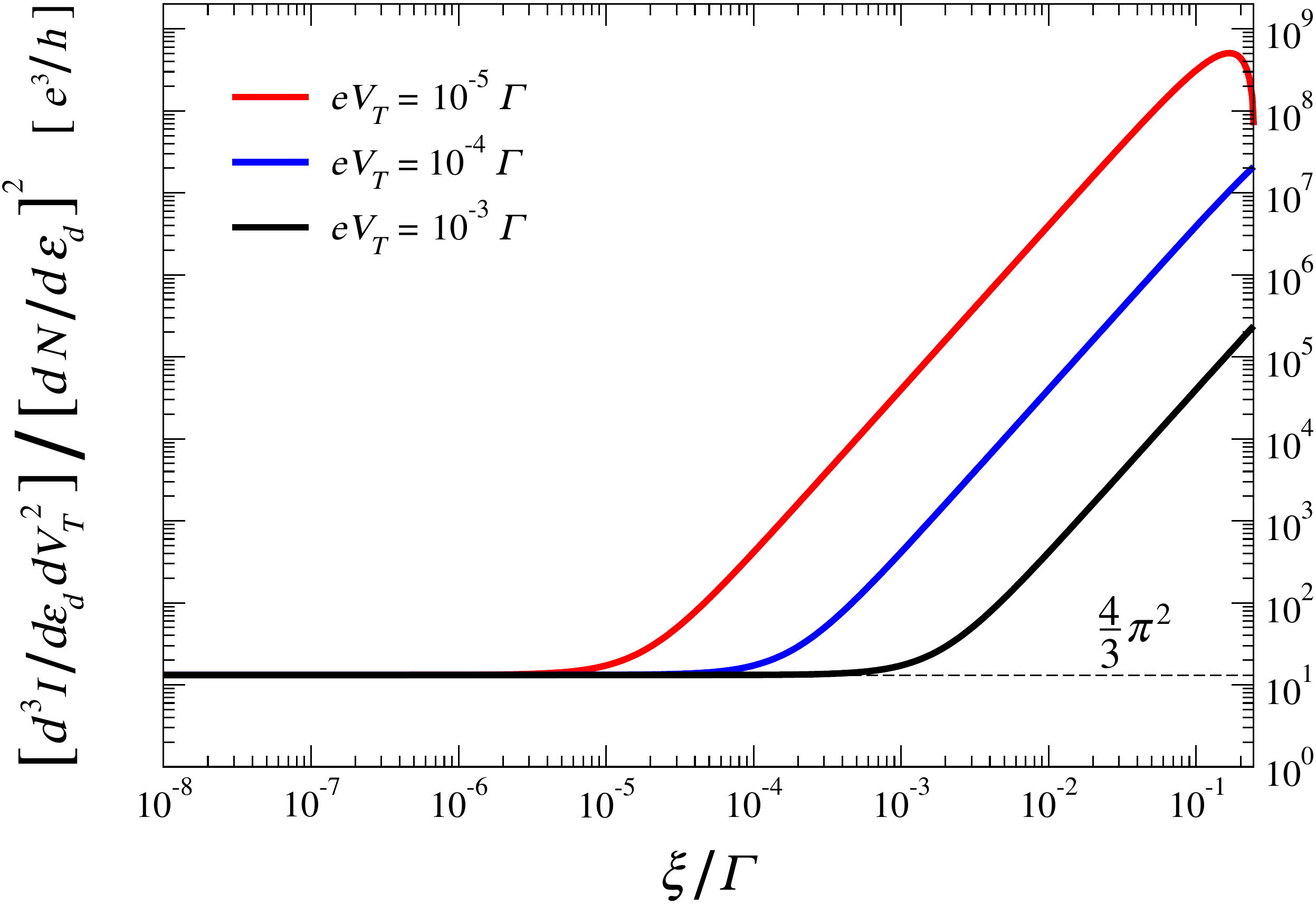}
\caption{\label{figure_6} The ratio between the derivative of the electric
  current
  $-\partial^3 I/\partial\mu\partial V_T^2=\partial^3 I/\partial\epsilon_d\partial V_T^2$
  and the square of the quantum dot compressibility
  $[\partial N/\partial\mu]^2=[\partial N/\partial\epsilon_d]^2$. Because of
  the logarithmic scale we invert the ratio's sign as compared
  to Fig. \ref{figure_5}. The ratio is shown as a function of the Majorana
  overlap energy $\xi$ for various values of the thermal voltage $V_T$. The
  other parameters have the same values as in Fig. \ref{figure_2}.}
\end{figure}
reasonable to explore what happens with the ratio
$[\partial^3 I/\partial\mu\partial V_T^2]/[\partial N/\partial\mu]^2$ when
$\xi$ is large enough. In the present context it is clear physically that
large $\xi$ means that $\xi>eV_T$. In this case the coupling between the two
Majorana zero modes is so strong that the second Majorana mode $\gamma_2$
quickly adjusts to the dynamics of the first Majorana mode $\gamma_1$ directly
reacting to variations of $V_T$. So that in response to the thermal voltage
$V_T$ the two Majoranas behave together as a single state which represents a
partially separated Andreev bound state. In contrast, if $\xi<eV_T$, the
second Majorana mode $\gamma_2$ does not follow $\gamma_1$ involved in the
dynamics induced by variations of the thermal voltage $V_T$ and nonequilibrium
states emerging in this case are of unique Majorana nature. This is, indeed,
what we see in Fig. \ref{figure_6} showing the ratio
$[\partial^3 I/\partial\mu\partial V_T^2]/[\partial N/\partial\mu]^2$
as a function of the Majorana overlap energy $\xi$ for three different values
of the thermal voltage $eV_T$. As one can see, when $\xi<eV_T$ and
nonequilibrium states result from fully unpaired Majorana zero modes, the
ratio is independent of $\xi$ and is equal to the universal Majorana value,
$[\partial^3 I/\partial\mu\partial V_T^2]/[\partial N/\partial\mu]^2=-4\pi^2e^3/3h$.
However, when $\xi\approx eV_T$, the ratio starts to deviate from the
universal Majorana plateau $-4\pi^2e^3/3h$. At this point both Majorana modes
$\gamma_1$ and $\gamma_2$ start to feel the variations of the thermal voltage
$V_T$ and respond to these variations together as a partially separated
Andreev bound state. As a result,
$[\partial^3 I/\partial\mu\partial V_T^2]/[\partial N/\partial\mu]^2$
significantly deviates from the universal Majorana value $-4\pi^2e^3/3h$
when $\xi$ is further increased as can be seen in Fig. \ref{figure_6} for
$\xi>eV_T$.
\begin{figure}
\includegraphics[width=8.0 cm]{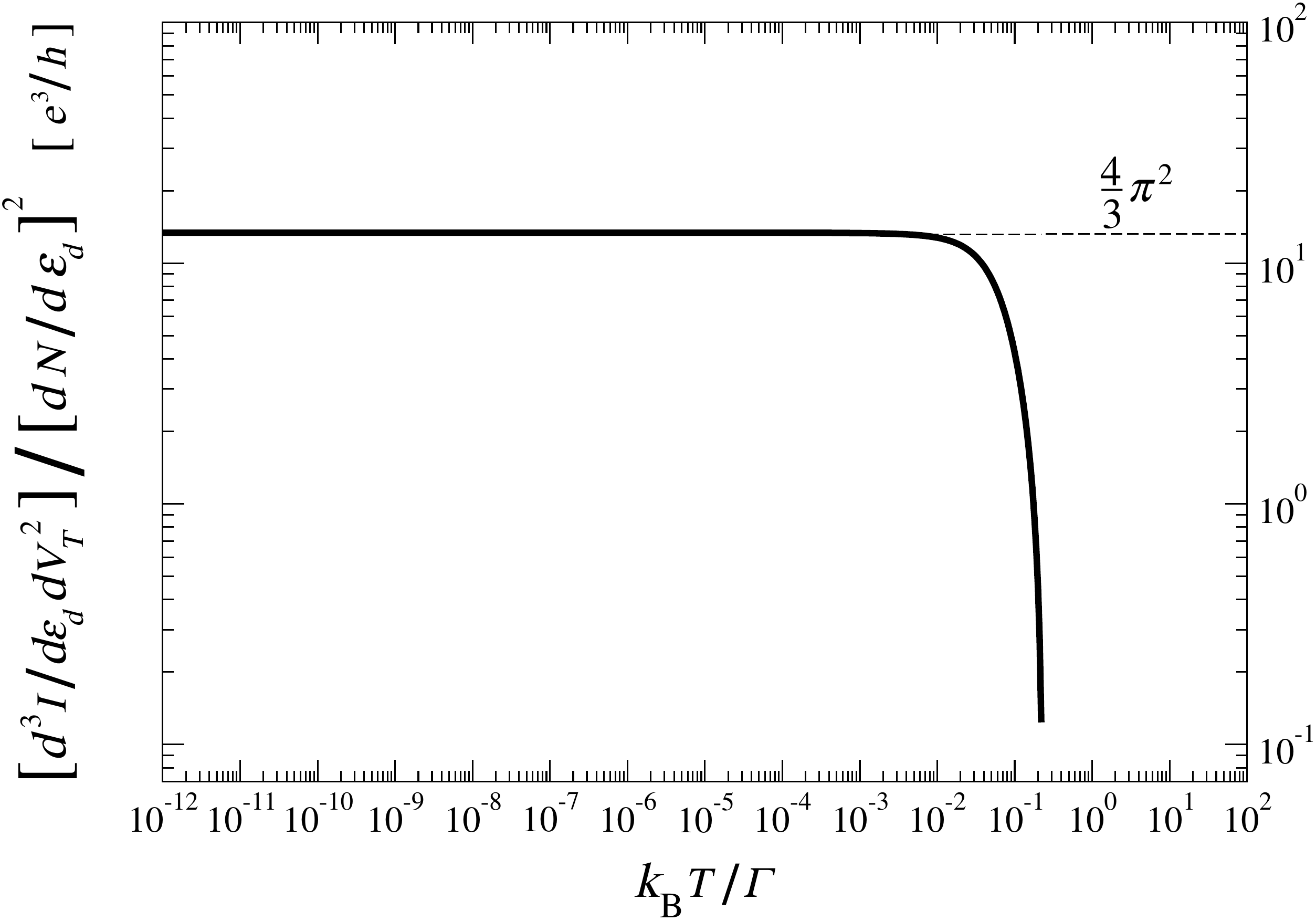}
\caption{\label{figure_7} The ratio between the derivative of the electric
  current
  $-\partial^3 I/\partial\mu\partial V_T^2=\partial^3 I/\partial\epsilon_d\partial V_T^2$
  and the square of the quantum dot compressibility
  $[\partial N/\partial\mu]^2=[\partial N/\partial\epsilon_d]^2$. As in
  Fig. \ref{figure_6}, because of the logarithmic scale we invert the ratio's
  sign as compared to Fig. \ref{figure_5}. The ratio is shown as a function of
  the temperature $T$. The thermal voltage is relatively large,
  $eV_T/\Gamma=10^{-2}$. Here $\epsilon_d/\Gamma=0.5$, $|\eta|/\Gamma=60$ and
  $\xi/\Gamma=10^{-8}$.}
\end{figure}

The above analysis has been done for almost zero temperature,
$k_\text{B}T/\Gamma=10^{-12}$. Nevertheless, below we demonstrate that the
dual Majorana universality analyzed above is robust against high temperatures
and high thermal voltages. In Fig. \ref{figure_7} the ratio
$[\partial^3 I/\partial\mu\partial V_T^2]/[\partial N/\partial\mu]^2$ is shown
as a function of the temperature $k_\text{B}T$ for the thermal voltage
$eV_T/\Gamma=10^{-2}$. The Majorana overlap energy here is small,
$\xi/\Gamma=10^{-8}$, so that $\xi\ll eV_T$ and, as discussed above,
nonequilibrium states clearly reveal unique Majorana physics. From the curve
shown in Fig. \ref{figure_7} one can see that the ratio has the universal
Majorana value $-4\pi^2e^3/3h$ for temperatures
$k_\text{B}T\lesssim eV_T$. For higher temperatures, $k_\text{B}T>eV_T$, the
ratio strongly deviates from the universal Majorana plateau. This implies that
for very high temperatures the Majorana zero modes are no longer effective in
formation of nonequilibrium states of purely thermal nature. However, the
temperature $k_\text{B}T/\Gamma=10^{-2}$ should be already high enough to be
reached in modern experiments. Indeed, the largest energy scale in
Fig. \ref{figure_7} is given by the energy $|\eta|$. This means that this
energy should not exceed the induced superconducting gap $\Delta$ and, for
example, one can take $|\eta|\sim\Delta$. Experiments in Ref. \cite{Wang_2013}
demonstrate that sufficiently high values, such as $\Delta\approx 15$ meV,
have already been achieved and may soon become regular for generating Majorana
zero modes. Since in Fig. \ref{figure_7} for $k_\text{B}T/\Gamma=10^{-2}$ we
have $k_\text{B}T=1.7\times10^{-4}|\eta|=1.7\times10^{-4}\Delta$, the
temperature at which one observes the universal Majorana value
$[\partial^3 I/\partial\mu\partial V_T^2]/[\partial N/\partial\mu]^2=-4\pi^2e^3/3h$
is estimated as $T\approx 30$ mK in the SI units. Such temperatures are
already high enough to be reached in modern experiments.
\section{Conclusion}\label{concl}
In conclusion, we have revealed Majorana universality originating
simultaneously from both transport and thermodynamic properties emerging in
nonequilibrium states of purely thermal nature. Combining two essentially
different physical properties of quantum dots, the current response and
compressibility, one may observe their universal correlation as highly unique
Majorana physics inaccessible within strictly transport or thermodynamic
experiments. Although this special kind of Majorana universality requires
measurements of both the current and compressibility, it is exactly this
combination which makes it much more unique than measurements of only the
current aiming to obtain the differential conductance. Importantly, since such
measurements involve the mean current, they are much simpler than measurements
of the current noise. Another possible benefit is that compressibility
measurements \cite{Martin_2008} involving single-electron transistors indicate
that this physical quantity may soon be accessed also in quantum dot
experiments and one expects that the complexity of such experiments will not
be higher than the one related to the thermodynamic measurements
\cite{Sela_2019} of the universal Majorana entropy \cite{Smirnov_2015}. The
above advantages and practical accessibility of the universal range of the
model parameters demonstrate that the previously unknown Majorana universality
predicted here is a feasible goal for state-of-the-art experiments on
nanoscopic Majorana setups.
\section*{Acknowledgments}
The author thanks Milena Grifoni, Andreas K. H{\"u}ttel and Wataru Izumida for
useful discussions.

\end{document}